An alternative view on dissipation in turbulent flows

Gilbert Zalczer

SPEC, CEA, CNRS, Université Paris Saclay, CEA Saclay,  91191 Gif sur Yvette France

Abstract :

An original experimental setup has been elaborated in order to get a better view of turbulent flows in a von Karman geometry. The availability of a very fast camera allowed to follow in time the evolution of the flows. A surprising finding is that the development of smaller whorls ceases earlier than expected and the aspect of the flows remains the same above Reynolds number of a few thousand. This fact provides an explanation of the constancy of the reduced dissipation in the same range without the need of singularity. Its cause could be in relation with the same type of behavior observed in a rotating frame, the largest whorls taking the role of the external rotation.

Introduction

Turbulent flows have always been a subject of fascination. The first attempt to rationalization is probably due to Richardson [1] who observed that the largest whorls gave rise to smaller ones and extrapolated that to very small scales and introduced the concept of a cascade. This cascade was quantified by Kolmogorov [2] who predicted an energy distribution at the different scales obeying a -5/3 power law, which was then experimentally observed. An intriguing point was the behavior of the dissipation which remained finite at very high Reynolds numbers while it was supposed to vanish [3]. This led Onsager [4] to conjecture the existence of singularities at small scale in the velocity flows.

Owing to recent technical progresses in imaging devices, a more direct insight in fast moving flows was made possible.

The experiment

Von Karman flows have been widely studied owing to their practical aspect. Indeed they allow flows with high Reynolds numbers in a compact and closed geometry. The fluid is in a cylindrical container with height and diameter approximately equal. It is moved by two contrarotating impellers at the top and the bottom. The geometry of the impeller can be varied. We used a system existing in the laboratory with dimensions about 0.1 m and curved blades. The cylindrical cell was immersed in a rectangular water-filled tank. A light sheet issued from a continuous wave solid state laser (445nm or 532 nm) and vertically widened by a cylindrical lens passes through the meridian plane of the cell as shown in figure 1. The impellers were directly driven by stepping motors which could be driven either in half step mode at higher speeds or in microstep mode for smooth motions at low speed.   The results we show were obtained with equal  speed and  "positive" rotation i.e. the curvature of the blades enhancing the centrifugal force. A camera images at right angle the suspended particles. We use a Phantom  M120 camera which allows a frame rate up to 1035 f/s at a resolution of 1460 x 1080 pixels. The images were recorded using the LaVision Davis software. The cell was filled with water for the high Re values. In order to reach Re values low enough to observe the laminar flow, we had to use solutions of either saccharose or glycerol at concentrations giving a viscosity about 10 times greater. The liquids were seeded with silver coated hollow glass beads of 100 µm diameter provided by La Vision. The frame rate was synchronized with the stepping clock of the motors (300 images per turn except for the highest speeds) and the exposure time set as long as possible. Settings of the lens aperture (1.4-16) and laser intensity (0.1 - 1 W) compensated for the exposure time. The series of images could be analyzed with usual methods, either using LaVision software or optical flow methods. However we were additionally able to see the time evolution of the flow. A much better visualization is obtained by watching the superposition of a few consecutive images (typically 10).

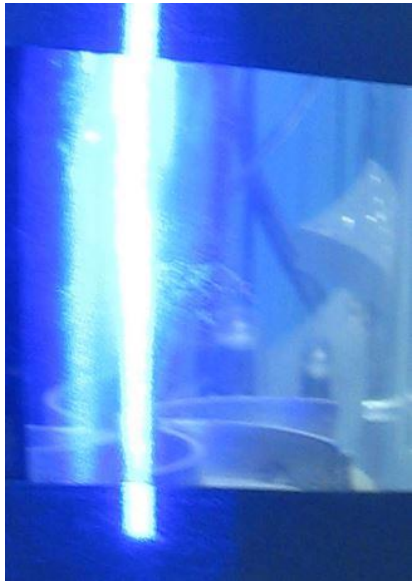

Figure 1 : Picture of the setup showing the laser plane and the lower impeller.

Observations

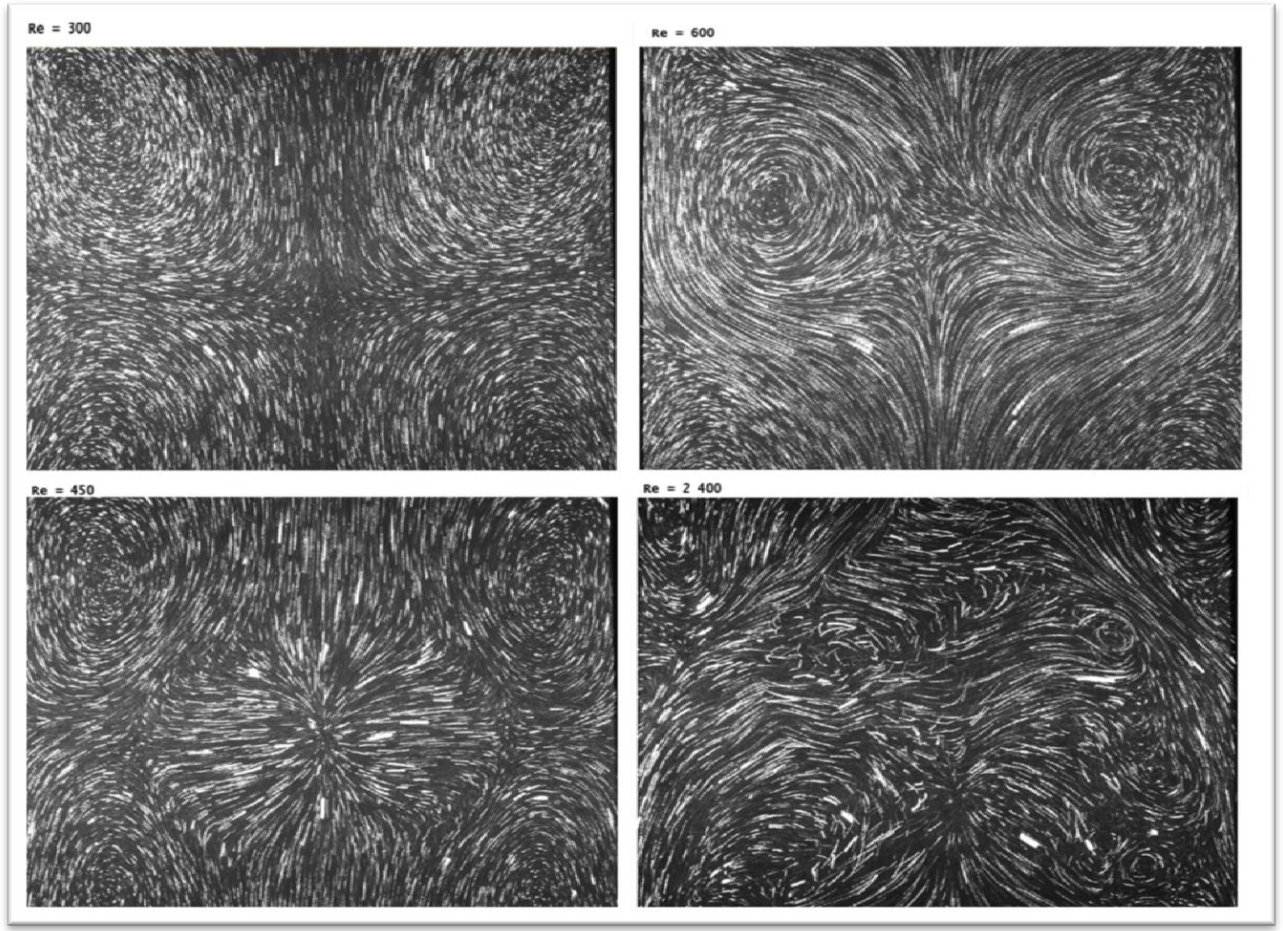

Figure 2 : Visualization of the flow at low Reynolds numbers. The base field is observed at Re=300, tpattern gets more complicated at 450 and 600 and many whorls can be seen at 2400.

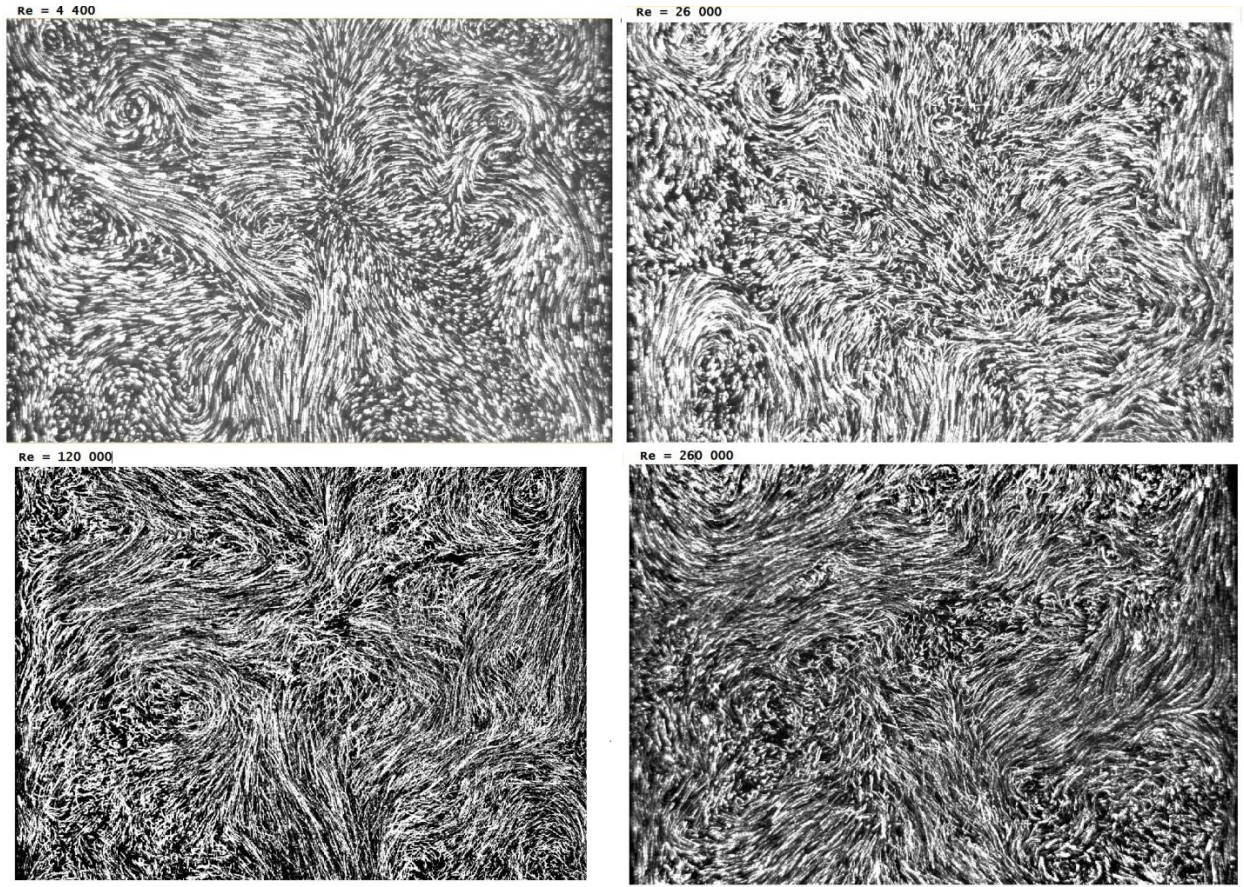

Figure 3 : Synthetic long exposure snapshots of flows at higher Reynolds numbers : above a few thousand, the flows cannot be distinguished.

A sample of images is reported in figures 2 and 3 for increasing Reynolds numbers. The base laminar flow gives way to more complicated ones with increasing Re. Following the words of Richardson, "big whorls have little whorls" when Re increases up to a few thousands ( In a limited region above this threshold the time evolution of the structures appearing after the destabilization of the base flow is slow so that some observed patterns can be metastable). However, beyond this value, no obvious smaller whorls appear and the overall aspect of the flow remains the same. This is still more flagrant when the time evolution can be watched on videos (http://www.dailymotion.com/gz_1). It has been briefly verified that Fourier transform of calculated velocity fields are consistent with the Kolmogorov prediction of a power law decrease with exponent -5/3 both in space and in time.

Discussion

An immediate consequence of this invariance is that the related dimensionless energy dissipation rate should be constant. It provides a very straightforward reason for this observed phenomenon. Indeed this experimental fact opens an alternative to the need of singular behavior of the flow at small scales.

The reason for this saturation of the Richardson cascade is yet to be determined. Cases where the turbulence does not fully develops are known. The first happens when the flow is confined in two dimensions. It has been shown [5] that in this case the cascade reverts and the energy flows from smaller to larger scales. The second is when the system under study is subjected to a global rotation. Indeed when this rotation is fast, it can be shown that the problem becomes equivalent to a two-dimensional one [6]. With more moderate rotation speeds the computation has not been yet been attempted but several experiments [7,8] showed a weakening of the turbulence intensity in line with a "two-dimensionalization" of the flow. It seems therefore a reasonable hypothesis that the large scale whorls induce the same effect as an external rotation and impair the smaller ones. In other words, an element of fluid which "feels" a rotation cannot distinguish whether it arises from a global rotation or a large whorl.

Due to technical difficulties, the experimental evidences are scarce on this subject, and it was admitted that singularities could be at most rare events. A recent experimental study [9], using very sophisticated methods, has detected possible footprints of singularities. However only 28 possible events were detected among 30000 samples while about 300 artefacts were discarded. The question whether these can account for the finite dissipation remains open. Only additional experiments, involving other geometries, together with new theoretical work will be necessary to settle the question.

Acknowledgements

I am particularly grateful to Antoine Seguin who lent me the Phantom camera at the center of this experimental study. Enric Meinhardt Llopis helped me a lot for image processing. I also thank Judith Vatteville for her help in using LaVision software, Darius Marin for his participation in the data gathering and Vincent Padilla for technical assistance.